\documentclass[twocolumn,prl]{revtex4}
\usepackage{graphicx}
\usepackage{dcolumn}% Align table columns on decimal point
\usepackage{bm}% bold math
\usepackage{amsmath}
\usepackage{times}
\usepackage{epstopdf}
\usepackage{color}
\usepackage[breaklinks=true,colorlinks,citecolor=blue,linkcolor=blue,urlcolor=blue]{hyperref}

\makeatletter

\newcommand{\Rmnum}[1]{\expandafter\@slowromancap\romannumeral #1@}
\makeatother

\begin{document}

\title{Negative longitudinal magnetoresistance in the density wave phase of Y$_2$Ir$_2$O$_7$}
\author{Abhishek Juyal}
\email{juyal@iitk.ac.in}
\author{Amit Agarwal}
\email{amitag@iitk.ac.in}
\author{Soumik Mukhopadhyay}
\email{soumikm@iitk.ac.in}
\affiliation{Department of Physics, Indian Institute of Technology Kanpur, Kanpur 208016, India}

\begin{abstract}
The ground state of nanowires of single crystalline Pyrochlore Y$_2$Ir$_2$O$_7$ is a density wave.
Application of a {\it transverse} magnetic field increases the threshold electric field for the collective de-pinning of the density wave state at low temperature, leading to colossal magnetoresistance for voltages around the de-pinning threshold. This is in striking contrast to the case where even a vanishingly small {\it longitudinal} magnetic field sharply reduces the de-pinning threshold voltage resulting in {\it negative}  magnetoresistance. Ruling out several other possibilities we argue that this phenomenon is likely to be a consequence of the chiral anomaly in the gapped out Weyl semimetal phase in Y$_2$Ir$_2$O$_7$.
\end{abstract}

\maketitle
With comparable energy scales of the spin-orbit coupling, the
effective band width, and the Coulomb correlation, Pyrochlore Iridates R$_2$Ir$_2$O$_7$ (R = Yttrium, or lanthanide elements) have emerged as a very promising quantum many body system. This has opened up the possibility of realization of a variety of novel quantum phases~\cite{Kim, Pesin, Wan, Yang, Kargarian, Machida} including Weyl semimetals~\cite{Wan, William, Go} in Pyrochlore Iridates. However till date experimental evidence of Weyl semimetal (WSM) in Iridates has remained elusive, except for a suggestion of metal-semimetal transition in the optical conductivity study of Eu$_2$Ir$_2$O$_7$~\cite{Sushkov}. Apart from the practical problem of fabricating clean single crystals of Iridates, other fundamental reason could be that Weyl nodes  gap out due to the predicted density wave instabilities caused either by coulomb correlation or induced by magnetic field~\cite{vivek, Axion-CDW, RN_CDW_Weyl, Bitan, Yang, Redell}.

In a recent work, we demonstrated the existence of charge density wave (CDW) order in single crystalline nanowires of Pyrochlore Iridates \cite{Abhishek}: Y$_{2-x}$Bi$_x$Ir$_2$O$_7$ with $x= 0$ (YIO) and $0.3$ (YBIO). However, the physical origin of the observed CDW instability in YIO and YBIO is not clear. It can arise from the natural propensity towards Peierls instability in low dimensional systems. More interestingly it can also be a consequence of spontaneously broken chiral symmetry in three dimensional WSM -- in which case it may carry signature of axion dynamics \cite{Axion-CDW, Kaiyuyang}. To address this issue, investigating the properties of the density wave state in presence of external electromagnetic field would be an effective way of either establishing or ruling out the WSM phase in Y$_2$Ir$_2$O$_7$.

In this letter, we show the existence of a CDW ground state in single-crystalline YIO nanowires, and study the transverse (${\bf H} \perp {\bf E}$) and longitudinal (${\bf H} \parallel {\bf E}$) magnetoresistance in the sliding state of the density wave. We show that the threshold electric field for the collective de-pinning of the density wave increases on application of a transverse magnetic field, and since the nonlinear conductivity, contributed by by sliding CDW, is much higher than that of the normal quasiparticles, this leads to a {\it colossal} magnetoresistance (MR). On the other hand, application of even a tiny  longitudinal magnetic field component leads to a significant reduction of the de-pinning threshold, resulting in {\it negative} MR. Among the possible factors leading to the observed negative MR, we rule out weak localization, current jetting, and the reconstruction of the Fermi surface in presence of a magnetic field. We suggest that this seemingly non-perturbative effect of a tiny magnetic field (${\bf H} \parallel {\bf E}$) on the de-pinning threshold and the observed negative MR may be a consequence of the chiral anomaly or the coupling of the phase fluctuations of the axionic CDW to the ${\bf E}\cdot{\bf H}$ term~\cite{Axion-CDW}.
\begin{figure}
\includegraphics[width=1\linewidth]{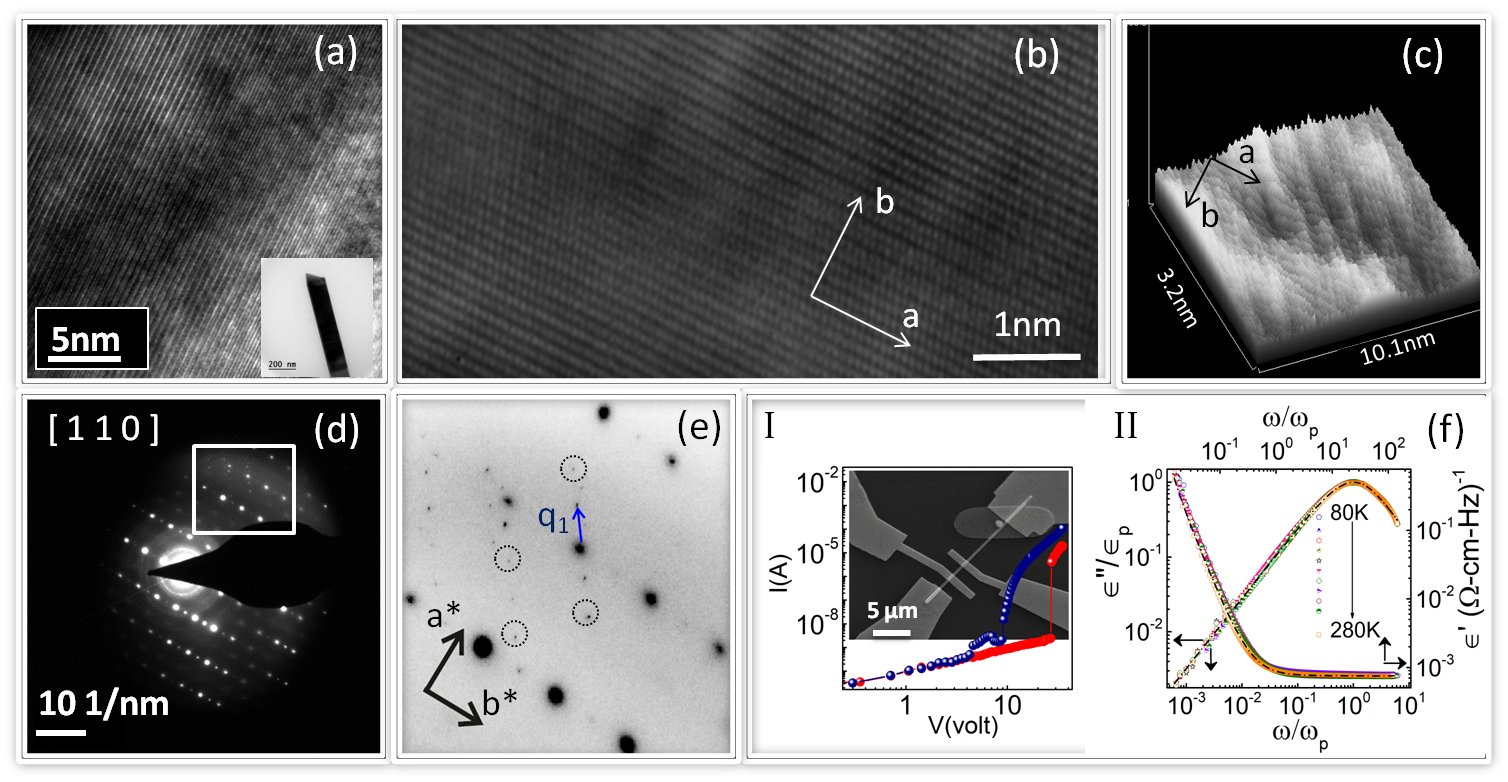}
\caption{(a) Real space HRTEM image of a portion of a 150 nm (diameter) YIO nanowire showing lattice fringes. Inset: TEM image of the same. (b) Magnified HRTEM image shows the periodic lattice distortion along the long axis `a' of the nanowire. (c) 3D projection of the real space image in (b) gives direct evidence of the periodic lattice distortion along `a' axis. (d) SAED pattern of the YIO nanowire in the (110) zone axis with distinct Bragg spots. (e) The area highlighted in (d) is zoomed in to show satellite spots (circled for improved visibility) due to density wave formation. (f) Panel I: DC IV characteristics of a switching/hysteretic (red curve) and a non-switching/non-hysteretic (blue curve) YIO device at $1.3$ K.
Inset: SEM image of a representative device for transport measurement. (f) Panel II: Dielectric spectroscopy of YIO: frequency dependence of the real and imaginary part of the dielectric permittivity, at different temperatures from 80K to 290 K scaled on a universal relaxation curve - Eq.~(\ref{acfit}) - represented by the solid red line. The temperature dependence of $\omega_{\rm p}$ (prominent relaxation frequency in the ac response) is given in the supplement~\cite{SM}.
\label{fig:tem}}
\end{figure}

Single crystalline nanowires of YIO were prepared following the method of \cite{Vinod, Abhishek}, and drop casted on silicon oxide substrate with subsequent metallization using e-beam lithography - see Fig.~\ref{fig:tem}(f).
For details of device fabrication and chemical  characterization, see~\cite{SM}.
High resolution transmission electron microscopy (HRTEM) and selected area electron diffraction (SAED) are sensitive to the periodic lattice distortions accompanying CDW. The SAED pattern in the (110) zone axis confirms that the lattice symmetry of the bulk is preserved in the nanowire (see Fig.~\ref{fig:tem}(d) and ~\cite{SM} for details). The real space HRTEM image in Fig.~\ref{fig:tem}(a) is characterized by a bright-dark contrast modulation due to the periodic lattice distortion. We also observe one-dimensional satellite spots confirming the presence of a new lattice periodicity which could be attributed to the CDW instability [Fig.~\ref{fig:tem}(e)]. A distinct and unique off-axis CDW structure can be seen in the SAED. The off-axis CDW shows a lattice modulation characterized by the wave vector ${\bf q}_1$ defined as ${\bf q}_{1} = 0.23 {\bf a}^{\ast} + 0.27 {\bf b}^{\ast}$ where $|{\bf a}^*|= 5.61/$nm and $|{\bf b}^*| =3.96/$nm [see Fig.~\ref{fig:tem}(e)]. The superlattice spots in other wave vector directions are not observed, thus negating any possibility of occurrence of `chiral CDW'~\cite{Ishioka} arising out of interference between coexisting density waves with separate wave vectors. The real space 2D HRTEM image at higher magnification and its 3D projection in Fig.~\ref{fig:tem}(b)-(c) provide direct visualization of the periodic lattice distortion.

Based on the low temperature IV characteristics (measured using pulsed IV method to eliminate self-heating effects) of a number of YIO nanowire devices we categorize them into switching and non-switching ones. For switching crystals, we find ohmic behaviour up to certain threshold voltage value $V_{\rm T}$, and beyond $V_{\rm T}$ there is a sharp hysteretic jump into a highly conducting non-linear state.
At higher temperature, the hysteresis and switching behaviour disappears, though there is still significant non-linear transport. For non-switching crystals we find highly nonlinear but smooth behaviour beyond the ohmic regime even at low temperature [Fig.~\ref{fig:tem}(f)].
The origin of nonlinear transport including hysteretic switching in CDW has been discussed earlier \cite{Vinokur, Levy, Littlewood, Hall, Strogatz}. The non-linear conduction, the switching behaviour and the associated threshold $V_{\rm T}$ can be dramatically tuned by changing the cooling rate~\cite{Abhishek}.
On rare occasions, we observe colossal enhancement of conductivity over the entire temperature range, when the sample is cooled sufficiently rapidly (see Fig.~2 of \cite{SM}). Since the impurity concentration is unlikely to change with change in cooling rate, this phenomenon could be related to the formation of dislocations in the CDW which possibly act as axion strings \cite{Axion-CDW}. The chiral modes along the axion strings can carry dissipation-less current leading to high conduction channels.

\begin{figure}
\includegraphics[width=1\linewidth]{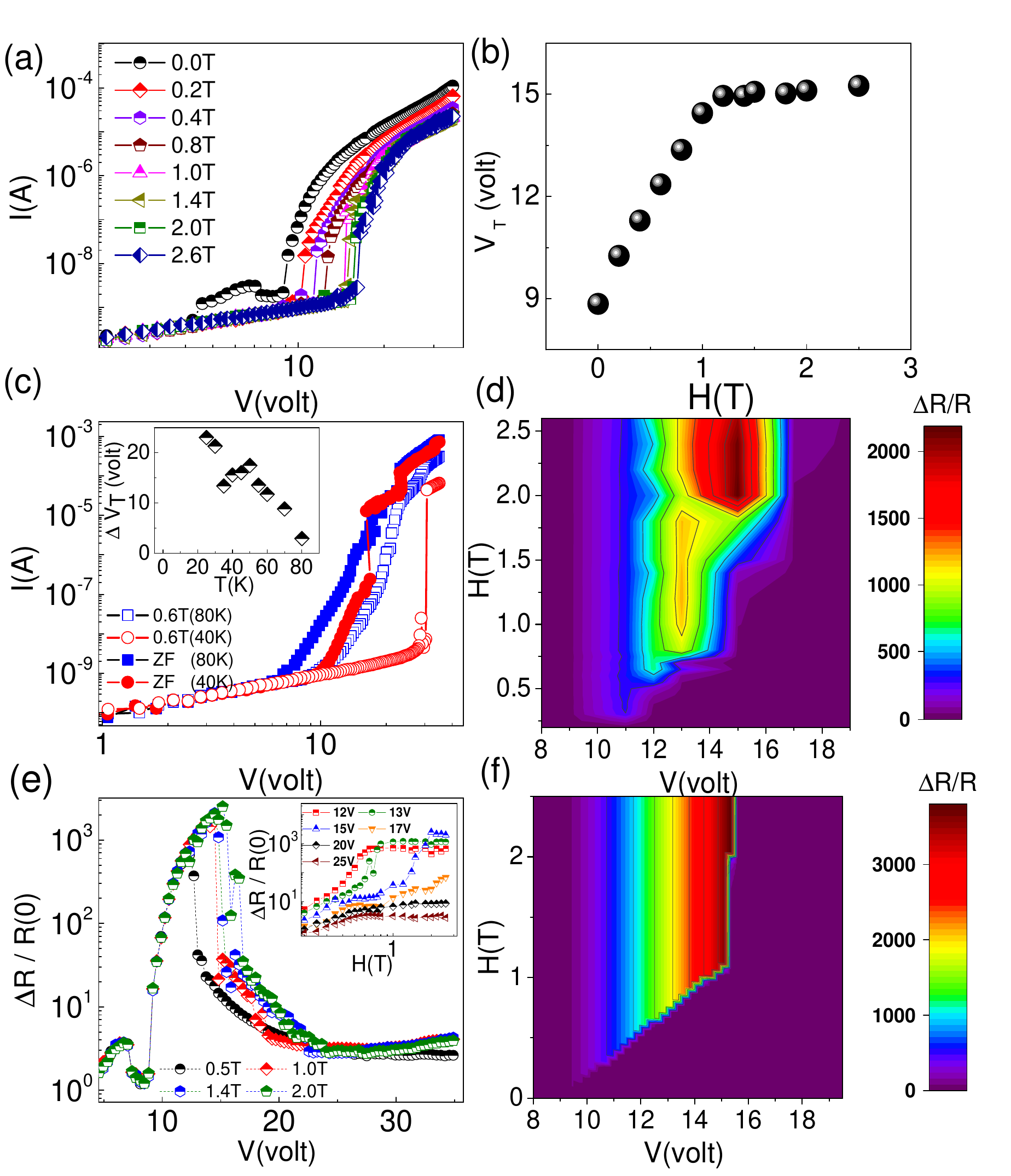}
\caption{(a) DC IV characteristics of a YIO nanowire (non-switching device) in presence of a transverse magnetic fields at 1.3K. (b) The magnetic field dependence of the threshold voltage $V_{\rm T}$. (c) The IV characteristics in presence of transverse magnetic field at higher temperatures for a switching device. At 40 K, the switching identified with discontinuity in the I-V characteristics for $H=0$ gets more pronounced in presence of transverse magnetic field. Inset: The temperature dependence of the change in $V_{\rm T}$ on application of $0.6$ Tesla magnetic field.
(d) 2D color plot of transverse MR in the $H-V$ plane for the same sample, which shows  {\it colossal MR} around the de-pinning voltage ($\sim 18$V). (e) The corresponding bias dependence of magnetoresistance (MR) at different magnetic fields.  The inset shows the $H$ dependence of MR for different bias voltages. (f) The broad qualitative features of the 2D color plot of the {\it transverse MR} in the $H-V$ plane are captured reasonably by a simple model including only the $H$ dependence of the threshold voltage $V_{\rm T}$.}\label{fig:outofplane}
\end{figure}

The low frequency dielectric response was measured at different temperatures, using the standard lock-in technique. In Fig.~\ref{fig:tem}(f), we show the real and imaginary part of the low frequency dielectric response (related to the ac conductivity via the relation $i \omega \epsilon(\omega) = \sigma(\omega) - \sigma_0$, where $\sigma_0$ denotes the dc conductivity) for several temperatures (from 80 K to 290K).
Similar dielectric response for YBIO nanowire was reported earlier~\cite{Abhishek}. The dielectric response at different temperatures [see Fig.~\ref{fig:tem}(f)] follows the generalized Debye relaxation formula \cite{Havriliak}, which includes a skewed distribution of the dielectric relaxation rates and is given by
\begin{equation} \label{acfit}
\frac{\epsilon(\omega) - \epsilon_{\infty}}{\epsilon_0 - \epsilon_{\infty}} = \frac{1}{\left[1 + (i \omega \tau(T))^{1-\alpha}\right]^\beta}~.
\end{equation}
Here $\alpha$ denotes the temperature dependent width, and $\beta$ the skewness of the distribution of dielectric relaxation rates.

The transport characteristics of single crystalline YIO nanowire device changes significantly in presence of a magnetic field. The dc electrical transport characteristics for 
non-switching as well as switching devices in presence of transverse magnetic field (${\bf H} \perp {\bf E}$) is shown in Figs.~\ref{fig:outofplane}(a) and (c), respectively.
We find that the threshold electric field, $V_{\rm T}$ increases with increasing transverse magnetic field and saturates  beyond field strength of $1$ Tesla [see Fig.~\ref{fig:outofplane}(b)]. Figure~\ref{fig:outofplane}(d), shows the transverse MR [$= R(H)/R(0)-1 = I(0)/I(H) -1$] as a function of the applied voltage and magnetic fields. Since there is linear (ohmic) conduction below $V_{\rm T}$, and nonlinear (CDW) conduction above it, the magnetic field induced shift in $V_{\rm T}$ results in huge enhancement of magnetoresistance ({\it colossal MR} upto $\sim 2 \times 10^3$) around the zero field de-pinning threshold  - as highlighted in Figs.~\ref{fig:outofplane}(d)-(e). The {\it colossal MR} is a consequence of the huge difference in the current carried by the normal carriers before the CDW starts sliding, and the large nonlinear current carried by the sliding CDW.

While the threshold voltage first increases and then saturates with increasing $H$ [Fig.~\ref{fig:outofplane}(b)], with increasing temperature, $\Delta V_{\rm T} = V_{\rm T}(H) - V_{\rm T}(0)$ decreases (almost linearly) as shown in the inset of Fig.~\ref{fig:outofplane}(c). The increase of threshold electric field in presence of transverse magnetic field in density waves has been addressed theoretically by Maki et. al.~\cite{Maki2}. In the weak pinning limit, a magnetic field perpendicular to the conducting plane of a DW reduces the elastic constant associated with the transverse phase distortion of the DW, leading to an increase in the threshold electric field. A similar argument for the increase of $V_{\rm T}$ with application of transverse magnetic field has also been shown to work in the strong pinning limit so long as there is imperfect nesting~\cite{Maki2}.

\begin{figure}
\includegraphics[width=1\linewidth]{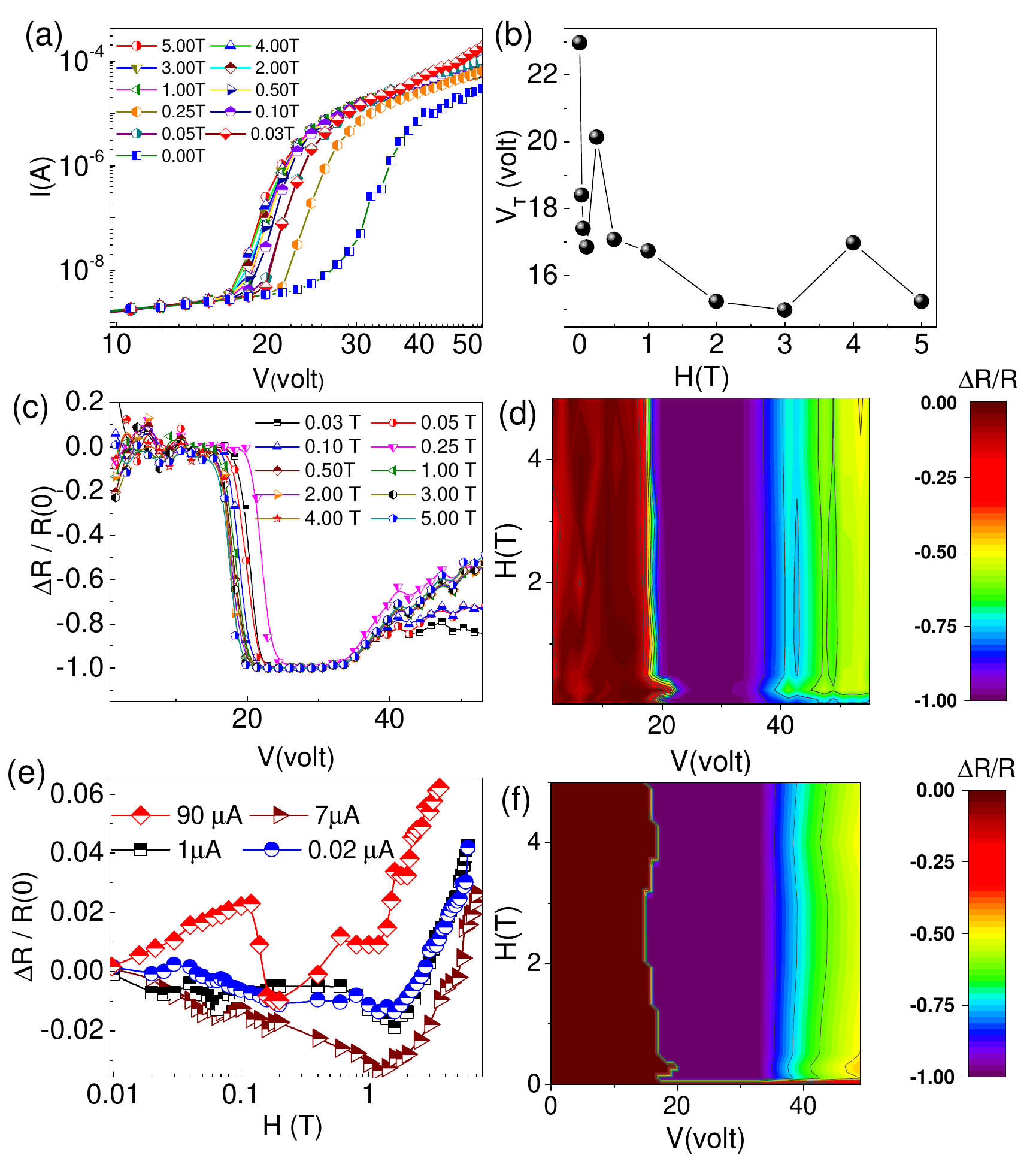}
\caption{(a) DC IV characteristics in presence of magnetic fields applied at $40^\circ$ to the electric field direction in a non-switching device.
(b) $V_{\rm T}$ abruptly shifts to lower value (from $23$ V to $17$ V) on application of even a small magnetic field.
(c) MR versus $V$ for different $H$ values. For $V \approx 18$ V we see the transition from almost negligible MR to significant negative MR, and for $V> 18V$ all curves have MR $<0$. The negative MR for $ V_{\rm T}(H)< V < V_{\rm T}(0)$ at all magnetic field values, is a consequence of the drastic reduction of $V_{\rm T}$ in presence of a longitudinal component of the magnetic field.
(d) 2D color plot of the experimentally observed MR in the $H-V$ plane.
(e) MR vs $H$ at constant bias currents showing negative MR at low magnetic field. (f) The broad qualitative features of the {\it longitudinal MR} in the $H-V$ plane is reasonably reproduced by the model discussed in the text.
}
\label{fig:inplane}
\end{figure}

The dc transport characteristics for low magnetic field change dramatically when a component of the magnetic field is applied along the current direction (in this case, the angle between {\bf{H}} and {\bf{E}} being $40^\circ$). The I-V curve for a non-switching YIO device is shown in Fig.~\ref{fig:inplane}(a). Focusing on the threshold voltage behaviour, even a small component of parallel magnetic field reduces $V_{\rm T}(H)$ drastically, in both strong pinning (switching samples -- see Fig.~{4} in ~\cite{SM}) and weak pinning (non-switching samples) regimes as shown in Fig.~\ref{fig:inplane}(b). On increasing the magnetic field further, $V_{\rm T}$ fluctuates a little with its mean value not changing significantly. This seemingly non-perturbative behaviour of sudden decrease of $V_{\rm T}$ even for a tiny magnetic field component along the transport direction, has not been reported earlier (to the best of our knowledge) in any CDW system. While a detailed theoretical understanding of this is lacking, there is a possibility of it being related to the chiral anomaly in Weyl semimetals, which leads to a charge
imbalance (proportional to ${\bf E}\cdot{\bf H}$)  between the two opposite chirality nodes.
This seems plausible, since even for a small parallel component of ${\bf H}$, the large value of the threshold voltage implies  that ${\bf E}\cdot{\bf H}$ is not insignificant.

The sudden reduction of $V_{\rm T}$ even for a tiny longitudinal ${\bf H}$ component, combined with the much higher CDW nonlinear conductivity as compared to the normal quasiparticles ohmic conductivity, leads to a negative MR in the bias range $V_{\rm T}(H) < V < V_{\rm T}(0)$.
The very high value of the CDW  nonlinear conductivity also limits the negative low field longitudinal MR [$= I(0)/I(H) -1$] to a minimum value of $-1$.

For bias fields above the zero magnetic field threshold, $V > V_{\rm T}(0)$, we observe positive MR similar to the case when the magnetic field is applied in the transverse direction. This simply indicates that the nonlinear sliding CDW conductance decreases with increasing magnetic field. {For devices which remain ohmic in zero magnetic field, we observe negative MR all throughout (see Fig. 3 (d) of \cite{SM} for an example).}

When the magnetoresistance is measured at a constant bias currents (instead of constant voltage), we observe negative MR at low magnetic field for low bias currents followed by positive MR at higher bias currents [see Fig.~\ref{fig:inplane}(e)]. This is consistent with the constant voltage measurements and can be understood qualitatively from Fig.~\ref{fig:inplane}(a): in a constant current measurement, increasing $H$ leads to decreasing $V$ for
iso-current condition. Thus the constant current MR ($= V (H)/V (0) - 1.0$) also has to be negative as long as the de-pinning current threshold $I_{T}(H)<I_{T}(0)$. The magnitude of the MR is however much less compared to that obtained for constant voltages. Usually, in non-switching samples there should be very little difference between constant voltage and constant current measurements. However, for switching samples or samples with strong non-linearity near the de-pinning threshold, the two measurements can differ widely~\cite{Sherwin}.

To highlight the importance of the variation of $V_{\rm T}(H)$ in the observed MR behaviour, we construct a simple phenomenological model. In the DW pinned regime, $V < V_{\rm T}(H)$, the quasiparticle current is ohmic and does not depend on the magnetic field, thus $I_{\rm qp}(H,V) = V/R$, where $R$ is the resistance and $V$ is the applied voltage across the nanowire device. In the sliding CDW regime, $V>V_{\rm T}(H)$, the current can be modeled via $I(H,V) = I_{\rm qp}(H,V) +  I_{\rm cdw}(H,V)$, where $I_{\rm cdw}$ is the nonlinear CDW current. We use a simple expression for the voltage dependence of CDW conductance based on Bardeen's tunneling theory \cite{Bardeen2}: $ I_{\rm dw}(V) = I_0 (V-V_{\rm T})e^{-V_0/V}$, where $I_0$ and $V_0$ are constants. % and $V_{\rm T} = E_{\rm T} L$, is the threshold voltage with $L$ being the nanowire length.
Incorporating the observed $V_{\rm T}(H)$ in the model described above, and making the very simple assumption of $I_0$ and $V_0$ being $H$ independent, the broad qualitative features of the MR observed for both the transverse as well as the longitudinal case could be reproduced, as shown in panel (f) of Fig.~\ref{fig:outofplane} and \ref{fig:inplane}, respectively.

There are several plausible reasons for the observed negative longitudinal magnetoresistance for ${\bf E} \parallel {\bf H}$ in our experiment: 1) YIO has been predicted to host Weyl fermions \cite{Wan} and in presence of electron-electron interactions Weyl semimetals have been shown to have propensity towards density wave instability \cite{vivek,Axion-CDW,RN_CDW_Weyl,Bitan}. The sliding CDW phase mode $\theta(t,x)$ can couple to external electromagnetic fields via an axion term which is proportional to $\theta {\bf E}\cdot {\bf H}$. Thus the so called chiral anomaly can lead to the observed $V_{\rm T}$ behaviour, and consequently induce a negative longitudinal magnetoresistance. This is consistent with the observed large shift in $V_{\rm T}$ even for a vanishingly small component of the applied magnetic field ${\bf H} \parallel {\bf E}$. 2) The `current jetting effect' where inhomogeneous current distribution within the voltage probe can lead to the measured voltage decreasing with increasing magnetic field -- an apparent negative longitudinal MR~\cite{Reis}. Systems prone to current jetting related artefact generally have a high field induced resistance anisotropy. Bulk single crystals of YIO are unavailable and among the single crystal Iridates available, only Nd$_2$Ir$_2$O$_7$ \cite{Tian} show sizable resistance anisotropy. 3) Weak localization of the `trivial' quasiparticles in the presence of impurities~\cite{Li} or quantum interference of CDWs~\cite{Inagaki}. 4) The magnetic field can cause a modification of the Fermi surface gap structure and change the ratio of quasiparticle and the CDW condensate density, thereby leading to lowering of $V_{\rm T}$ and negative magnetoresistance~\cite{Coleman}.

In the present case, current jetting can be ruled out with reasonable surety on account of the following reasons: (a) the contact electrodes in our case are not point contacts. They are e-beam deposited metal contacts covering the entire cross section of the nanowire. (b) the diameter of the nanowire in our devices are less than 200 nm with uniform cross-section, much less compared to the typical dimensions (few milimeters) of the samples showing current jetting \cite{Reis}. (c) the diameter to voltage probe length ratio of YIO nanowire is close to 15 and above, large enough to rule out inhomogeneous current distribution playing any part. (d) Finally, the observed negative MR is not `non-saturating', as is the case with current jetting effect~\cite{Reis}. At higher magnetic field, we observe distinct positive contribution to the longitudinal MR.

Weak localization of normal carriers can be ruled out due to the following reasons: a) there is no negative MR in presence of transverse magnetic field. (b) when the magnetic field is applied along the current direction, we do not observe any change of resistance corresponding to the ohmic regime in the IV characteristics. This completely rules out the role of the normal quasiparticles (not condensed into the DW) in the observed MR behaviour. Since there is no negative MR in presence of transverse magnetic field one can rule out negative MR arising out of quantum interference of CDWs as well. Additionally the applied magnetic field is too low to modify either the band gap or the condensate density of an ordinary CDW~\cite{Bjelis}. A rough estimate at low magnetic field of $H=0.01$T, for Fermi velocity $v_{F}=10^5$m/s and CDW order parameter $\Delta_{0}=300$K, gives the ratio of the cyclotron frequency $\omega_c$ in the CDW to $\Delta_0$ to be $\omega_{c}/\Delta_{0}\sim 0.01$.

In conclusion, the existence of CDW in single crystalline YIO nanowire is independently confirmed by HRTEM and SAED measurements. Additionally  we show that: 1) Application of a transverse magnetic field leads to an increase in the de-pinning threshold voltage in the CDW phase, leading to {\it colossal positive MR} which decreases with increasing temperature. 2) In presence of  small longitudinal magnetic field, the CDW sliding threshold voltage decreases significantly resulting in {\it negative MR}. These observations strongly suggest the role of finite  ${\bf E}\cdot {\bf H}$ dependent chiral anomaly at play in the dynamics of the DW state of single crystalline YIO nanowire.

\end{document}